\documentclass[proceedings]{JHEP3}

\PrHEP{PrHEP hep2001}                   
\conference{International Europhysics Conference on HEP}
\usepackage{epsfig}

\title{How Neutrino and Charged Fermion
Masses Are Connected Within Minimal Supersymmetric SO(10)}

\author{Borut Bajc\\
J. Stefan Institute, Ljubljana, Slovenia\\
E-mail: \email{Borut.Bajc@ijs.si}}                 
\author{Goran Senjanovi\'c\\
International Centre for Theoretical Physics, Trieste, Italy\\
E-mail: \email{goran@ictp.trieste.it}}             
\author{\speaker{Francesco Vissani}\\             
        INFN, Laboratori Nazionali del Gran Sasso, Theory Group, Italy\\
        E-mail: \email{vissani@lngs.infn.it}}     

\abstract{Massive neutrinos are a generic prediction of SO(10),
and models of unification cry for supersymmetry. Since we have 
a rather detailed information on neutrino and charged fermion masses, 
the real question is: how/whether it is possible to build a
SO(10) supersymmetric model, that correctly 
incorporates fermion masses. We show that 
a {\em simple construction} is possible in the context 
of a minimal theory.
We concentrate on the two heaviest generations, 
discuss the predictions of the model, and briefly 
comment on open questions.}

\begin{document}
\section{Yukawa Couplings at $M_{\sf GUT}$}
In order to avoid unacceptably big
Dirac neutrino masses in SO(10) \cite{gfm},
one introduces ${\bf \underline{126}}$-plets scalars. 
These produce huge Majorana masses for 
$\nu^c$ \cite{gsr}, and decouple them from 
the light spectrum:
\begin{equation}
{\cal L}=-{{\bf 16}_i}\left[
Y^{(10)}_{ij} {\underline{\bf 10}}\, +\, 
Y^{(126)}_{ij} {\underline{\bf 126}}\,  
\right] {{\bf 16}_j}+h.c.
\end{equation}
The ${\bf \underline{10}}$-plet contains two Higgs doublets,
that we call $\varphi_u$ and $\varphi_d,$ while the 
${\bf \underline{126}}$-plet
 contains one singlet $S$ (needed for $\nu^c$),
one triplet $\Delta$ (which may contribute to light 
neutrino masses \cite{ms})
and two doublets $\varphi_u'$ and $\varphi_d'$ 
(useful to make up for wrong SO(10) mass relations
\cite{bm}). Indeed, with a self-explanatory notation for the 
Weyl fermions \cite{barb}:
$$
\left\{
\begin{array}{l}
{\bf 16}_i\ {\underline{\bf 10}}\ {\bf 16}_j
\ni
{\varphi_u}\ ({u^c_i u_j} + {\nu^c_i \nu_j})+
{\varphi_d}\ ({d^c_i d_j} + {\rm e}^c_i {\rm e}_j) +\ (i\leftrightarrow j);\\
{\bf 16}_i \ {\underline{\bf 126}}\ {\bf 16}_j
\ni
\frac{1}{2}({S}\ {\nu^c_i \nu^c_j} 
 +{\Delta}\  {\nu_i \nu_j}) +
{\varphi_u'} ({u^c_i u_j} -{3} {\nu^c_i \nu_j})+
{\varphi_d'} ({d^c_i d_j}-{3}
{{\rm e}^c_i {\rm e}_j}) +\ (i\leftrightarrow j)
\end{array}
\right.
$$
\noindent In this work, we propose 
a model of the Yukawa couplings,
in which all the features of 
the minimal SO(10) theory are exploited.

\vskip-2mm
\section{Beyond the Great (Supersymmetric) Desert} 
The question of starting up 
model building is:
\underline{what does the minimal supersymmetric} 
\underline{standard model (MSSM) want from SO(10)?}
We get an answer by extrapolating
the Yukawa couplings from $T=0$ to $T=\log(M_{\sf GUT}/M_Z)/2\pi\approx 5.2$
(see appendix A for details). From 
figure 1, one sees that:
\begin{itemize}
\item For $3^{rd}$ family charged fermions masses:
the {\em Hypothesis} of leading ${\bf \underline{10}}-$plet Yukawa
coupling 
\cite{gn}, 
that gives {$y_t=y_b=y_\tau$} at $M_{\sf GUT}$ is OK.\footnote{We 
tuned the {\em vev} ratio $\tan\beta=\langle H_u\rangle / \langle H_d\rangle \sim 55.4$
to get this. We use 1 loop ``running'' and $\alpha_3=0.118.$}

\item For $2^{nd}$ family charged fermion masses: the
{\em Hypothesis} of leading ${\bf \underline{126}}-$plet 
Yukawa coupling \cite{gj}, 
that gives
{$y_\mu=-3\times y_s$} at $M_{\sf GUT}$ is OK. 
\end{itemize}

\noindent This could be an accidental fact, 
but is suggestive enough to take it seriously.

\section{Determining Model and Parameters}
Now that we defined the target, the question becomes:
\underline{how to match MSSM and SO(10)} 
\underline{Yukawa couplings?}
SO(10) can meet the MSSM needs (illustrated in previous figure) 
after the very simple identification of the MSSM Higgs fields:
$H_u\approx \varphi_u$
and $H_d\approx \varphi_d+\varepsilon\ \varphi'_d.$ 
(Of course, the orthogonal doublets should decouple 
from the MSSM spectrum, to maintain gauge coupling unification--
namely, we need a ``doublet-doublet'' splitting).

This position leads us to identify the MSSM Yukawa 
couplings in the following manner:
\begin{equation}
\left\{
\begin{array}{l}
Y_u\approx Y^{(10)} \ \ \ \  {\mbox{\em diagonal by definition}}\\
Y_d\approx Y^{(10)} + \epsilon\ Y^{(126)}\\
Y_e\approx Y^{(10)} -3\; \epsilon\ Y^{(126)}\\
\end{array}
\right.
\end{equation}
\noindent Since we know the Yukawa couplings 
(after extrapolation at $M_{\sf GUT}$), we can deduce 
the size of several elements of the SO(10) Yukawa matrices.
The chain of deduction we follow and 
the numerical values we obtain
at $M_{\sf GUT}$ are shown in this table:
\begin{center}
\begin{tabular}{|ccl|}
\hline
$y_t,y_b,y_\tau$&$ \Rightarrow$ &
{$Y^{(10)}_{33}\simeq 0.94 $} $\gg \epsilon\, Y^{(126)}_{33}$ \\
$y_\mu, y_s $&$\Rightarrow$ &
{$\epsilon\, Y^{(126)}_{22}\simeq 1.4\times 10^{-2}$} 
$> Y^{(10)}_{22}$ \\
$y_c $&$ \Rightarrow$ & {$Y^{(10)}_{22}\simeq 1.8\times 10^{-3}$} \\
$V_{cb}$&$\Rightarrow$ &
{$\epsilon\ Y^{(126)}_{23}\simeq 2.7\times 10^{-2}$}\\
\hline 
\end{tabular}
\end{center}

\newpage
\begin{figure}[thb]
\vskip-5mm
\begin{center}
{{\tiny.\,\,\,}}\includegraphics[width=4.1cm,angle=270]{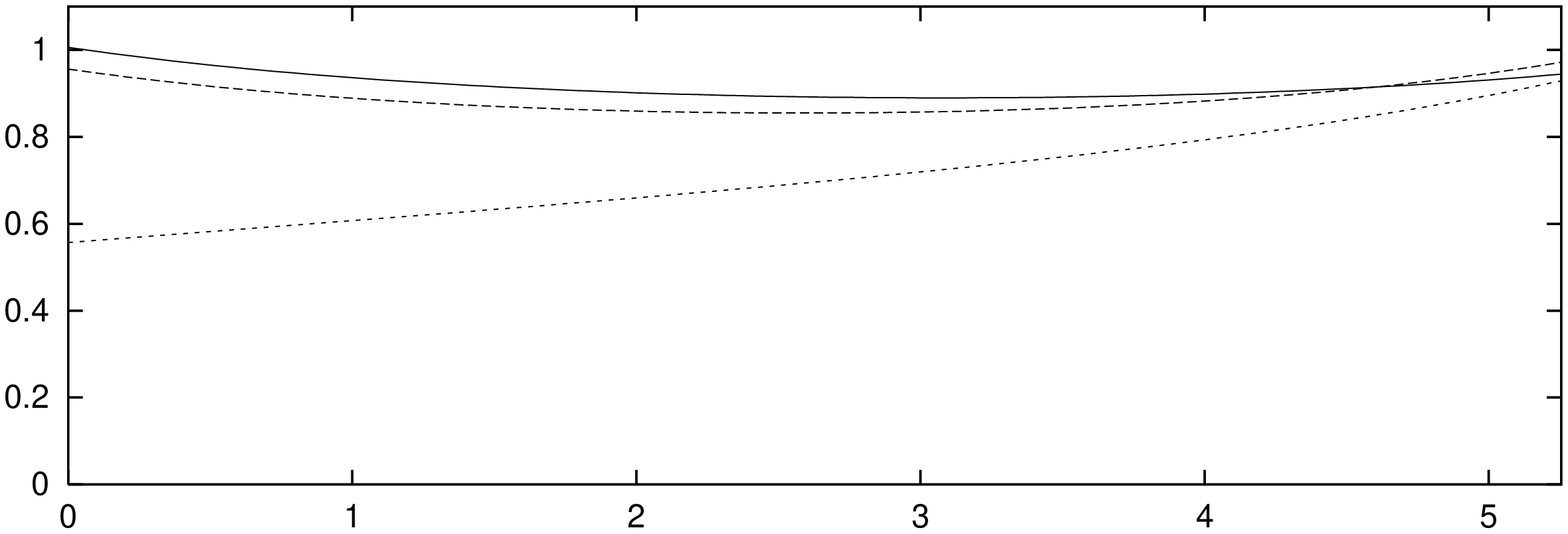} 
\vskip-3mm
\includegraphics[width=4.4cm,angle=270]{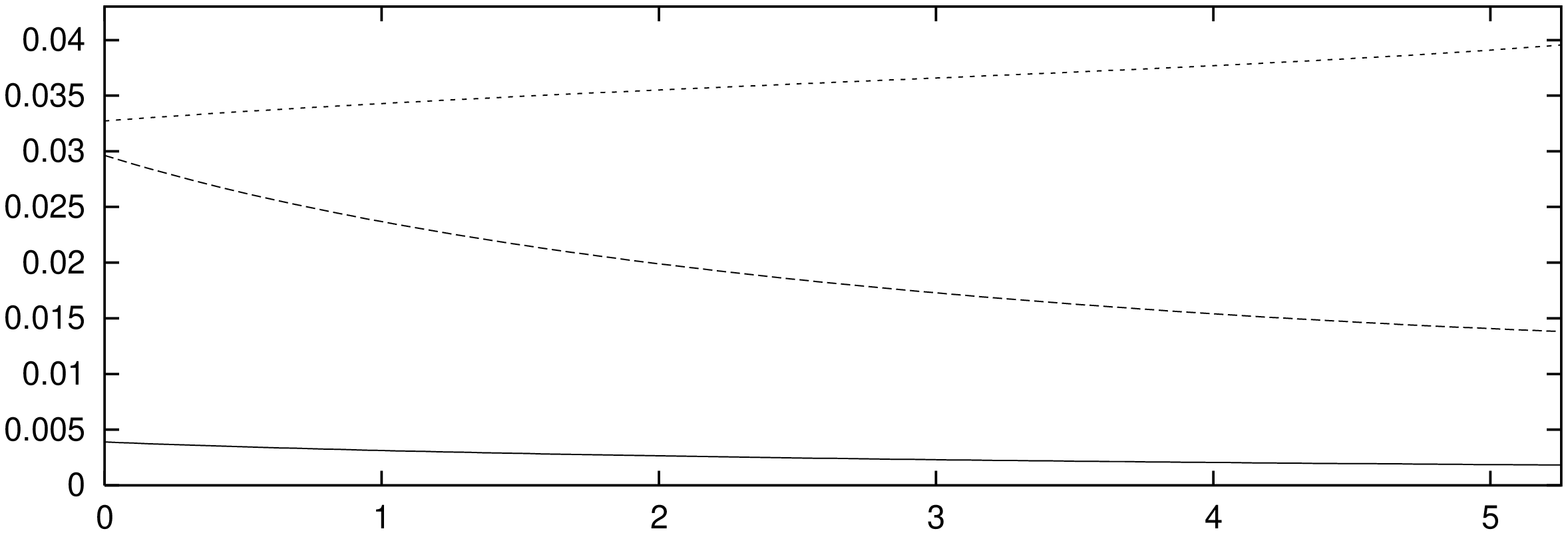}
\end{center}
\vskip-6mm
\rightline{$\log(Q/M_Z)/2\pi\ \ \ \ \ \ $}
\caption{Upper panel: Running of MSSM Yukawa couplings
of third generation from $M_Z$ till $M_{\sf GUT}$
($y_t$ is the largest at 
$M_Z$, $y_\tau$ the smallest). Lower panel: 
same for second generation ($y_\mu$ is 
the largest, {$y_c$} is the smallest). (We 
denote by $y_x$ the Yukawa coupling of the 
particle $x,$ e.g.: $y_t$ for top, $y_c$ for charm, 
$y_\mu$ for muon. For a given $\tan\beta,$
$y_x$ is computed from the mass of $x$ at $T=0.$)} 
\vskip2mm
\end{figure}

\noindent Two remarks are in order:
\begin{itemize}
\item We kept the deduction as simple as possible
{\em e.g.}\ we did not perform detailed diagonalizations 
to get these numbers, which saves us from considering their
phases. (However, we feel that it is 
fair to say that higher order effects, threshold and  
non-log corrections {\em etc.}\ could make a
much more accurate treatment meaningless.)
\item The only unknown element of the $2^{nd}-3^{rd}$ family blocks
is {$\epsilon\ Y^{(126)}_{33}$} (though one may 
reasonably guess that it is
not too far from {$\epsilon\ Y^{(126)}_{22}$}
or {$\epsilon\ Y^{(126)}_{23}$}).
\end{itemize}
Till here, we showed that the model is not contradicting known things...
\section{Neutrino Features}
Now we come to the fermion of the day: the neutrino.
In order to formulate our proposal, we will base our discussion on this
provocative question: \underline{what do these neutrinos want?}
We recapitulate the experimental situation by means 
of the  following table:
\begin{center}
\begin{tabular}{|clr|}
\hline
{$\Delta m^2_{31}$}  &  $[1.5,5]\times 10^{-3}$ eV$^2$   & {\sf atmospheric neutrinos} \\
{$\Delta m^2_{21}$}  &  $[2,50]\times 10^{-5}$ eV$^2$   & {\sf solar LMA \tiny (or $<2\times 10^{-7}$ eV$^2$)} \\
{$\theta_{23}$}  &  $[35^\circ,55^\circ]$   & {\sf atmospheric neutrinos} \\
$\theta_{13}$  &  $< 10^\circ$   & {\sf CHOOZ+atm.+K2K\ 
\tiny (depends on $\Delta m^2_{31}$)} \\
$\theta_{12}$  &  $[25^\circ,43^\circ]$   & {\sf solar neutrinos (99 \% CL)} \\
\hline
\end{tabular}
\end{center}
We will be mostly concerned with the first three items. As remarked 
by several people (see {\em e.g.}\ \cite{fv})
a neutrino mass matrix with a ``dominant block'' 
is strongly suggested:
$$
{
\frac{{\bf M}_\nu}{\sqrt{\Delta m^2_{31}}}=
\frac{1}{2}\times 
\left(
\begin{array}{ccc}
0 & 0 & 0 \\
0 & 1 & 1 \\
0 & 1 & 1 
\end{array}
\right)} +
{\cal O}\left(\theta_{13},\ \theta_{23}-\frac{\pi}{4},\ 
\sqrt{\frac{\Delta m^2_{21}}{\Delta m^2_{31}}} \right)
$$
But, due to hierarchical Yukawa couplings,
the seesaw does not yield this pattern {\em generically}
(however, see also \cite{go}).
Often, small values of $\theta_{23}$ are found, 
as pointed out in \cite{bm,biswa} and as 
illustrated here: 
$$
M_D M^{-1}_R M_D=
\left(
\begin{array}{cc}
\epsilon & 0 \\
0 & 1
\end{array}\right)
\cdot
\left(
\begin{array}{cc}
a & b \\
b & c
\end{array}\right)
\cdot
\left(
\begin{array}{cc}
\epsilon & 0 \\
0 & 1
\end{array}\right)
$$
Thus, we are lead to try another mass mechanism,
and we welcome the fact that we have the triplet $\Delta$
at our disposal  \cite{ms}
(by the way, we arrived at a common sense answer
to the question on ``neutrino wishes'':
\underline{neutrinos want to be different from the other fermions}).
\section{The Triplet Option}
We are assuming that neutrinos 
take mass {\em mostly} from the triplet $\Delta:$ 
${\bf M}_\nu \propto Y^{(126)}.$  
Running back to $M_Z$ the 
MSSM Yukawa couplings, we get a simple expression
for the $\nu_\mu-\nu_\tau$ block of the neutrino mass matrix:
\begin{equation}
{\bf M}_\nu \propto
\left(
\begin{array}{cc}
1 & 1.7 \\
1.7 & x
\end{array}
\right)
\label{e}
\end{equation}
(We have ``$x$'', for $Y^{(126)}_{33}$ is unknown, and also 
because seesaw {\em might} contribute to 33-entry---see {\em e.g.}
\cite{anjan}). Clearly, eq.\ (\ref{e}) can underlie a
``{dominant block}'', thus:
\vskip-4mm
\begin{center}
\fbox{{\sf $\theta_{23}$ can be large}}
\vskip2mm
\fbox{{\sf we expect a weak mass 
hierarchy (not $m_3\gg m_2$)}}
\end{center}
These two properties correlate, as can be seen 
in figure 2.
\begin{figure}
\begin{center}
\hskip13mm{\includegraphics[width=8cm]{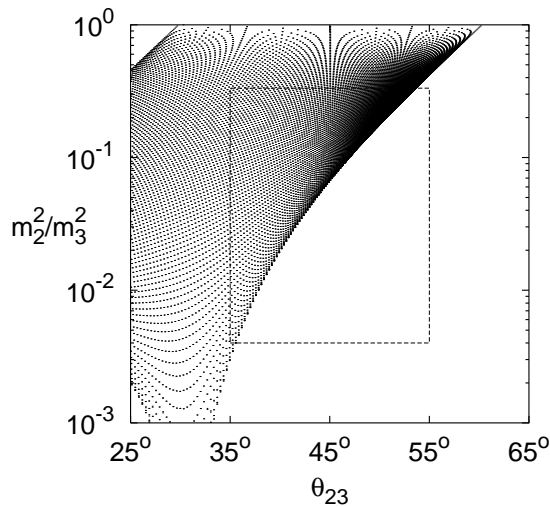}}
\end{center}
\caption{Possible values of the mass hierarchy parameter
$m_2^2/m_3^2$ and of the atmospheric mixing angle $\theta_{23},$
obtained varying the complex input parameter $x$
(eq.\ \protect\ref{e}). A rectangle encloses the range 
of permitted values, estimated assuming that the lightest neutrino mass
$m_1$ is negligible.} 
\end{figure}
To further illustrate this result
(assuming $m^2_3\ \simeq \Delta m^2_{31}=3\times 10^{-3}$ eV$^2$
and $m^2_2\ \simeq \Delta m^2_{21}$) 
 we note that:
\begin{center}
{$\bullet$} {If $\theta_{23}=45^\circ,$ then 
$\Delta m^2_{21}{>2\times 10^{-4}}$ eV$^2$;}

{$\bullet$} {If $\Delta m^2_{21}=5\times 10^{-5}$ eV$^2,$ then\footnote{Quite
tough to test experimentally, 
since it is equivalent to $\sin^22\theta_{23}<0.97...$} 
$\theta_{23}{<40^\circ}.$}
\end{center}

We conclude that the minimal SO(10) model for Yukawa coupling
we propose {\em is} predictive, despite (thanks to?) its simplicity.

\section{Summary and Discussion}
{\sf
\noindent {$\star$}
We discussed an ``economical
embedding'' of MSSM into SO(10), in a sense that
all features of {${\bf \underline{126}}$-plet}
have been exploited, namely: we use singlet, 
doublets {\it and} triplet $vev$'s.

\vskip2.0mm
\noindent {$\star$}
The most important step in the construction: how the masses
of the charged fermions of the {$2^{nd}$ and $3^{rd}$} 
generations are explained (Sects.\ 2 and 3).
$3^{rd}$ family unification suggests the
large $\tan\beta=\langle H_u\rangle / \langle H_d\rangle$
regime; this is not an appealing case, 
but perhaps it is still viable 
(incidentally, it permits us to accommodate a ``heavy'' Higgs
field, $m_h< 135$ GeV).

\vskip2.0mm
\noindent {$\star$}
The {\it triplet}
mechanism for 
neutrino mass generation is at least likely
(discussion in Sect.\ 4).
The correlations among
$({\bf M}_\nu)_{22}\leftrightarrow m_\mu, m_s,$ and
$({\bf M}_\nu)_{23} \leftrightarrow V_{cb}$ 
imply (eq.\ (\ref{e})):
\vskip-3.5mm
$$
{\theta_{23}}\in [\ 35^\circ\ ,\ 55^\circ\ ]
\Leftrightarrow 
{\frac{m_2^2}{m_3^2}}\in 
\left[\ \frac{1}{250}\ ,\ \frac{1}{3}\ \right]
$$
\vskip-1mm
\noindent Solar $\nu$ 
solutions with big hierarchy are disfavored, while 
LMA fits well the scheme.
After the $\Delta m^2_{21}$ measurement--at 
KamLAND?--we will get an upper bound 
on $\theta_{23}$ (fig.\ 2 and Sect.\ 5). 

\vskip2.0mm
\noindent {$\star$}
A {pending} question is:
masses of $1^{st}$ family fermions (also $m_1$);
proton decay rate; feasibility of 
baryogenesis-through-leptogenesis mechanism. 
These features are strictly tied
among them, and require further study.}
\vskip5mm
To conclude, we stress 
the main goals achieved: We showed that 
it is possible to build a simple
model for fermion masses based on 
supersymmetric SO(10), with renormalizable couplings only. 
This model accounts for the masses of second and
third generation fermions. It has large $\theta_{23},$
and prefers the solar neutrino solutions with weak mass hierarchy.

\vfill
\noindent{\large\bf Acknowledgments}
\vskip4mm
\noindent F.V.\ thanks the Organizers of the ``International 
Europhysics Conference on HEP'' for the beautiful conference, and the
Conveners for kind invitation and discussions. 
The work of B.B.\ is supported by the Ministry of Education, Science
and Sport of the Republic of Slovenia.
The work of G.S.\ is partially supported by EEC, under 
the TMR contracts ERBFMRX-CT960090 and HPRN-CT-2000-00152.
We express our gratitude to INFN, which 
permitted the development of the present study 
by supporting an exchange program with the 
International Centre for Theoretical Physics.

\appendix

\section{1 loop renormalization group equations}
We assume supersymmetry, in order to comply
with one-step unification of gauge couplings. 
The renormalization group equations relevant to our analysis are:
$$
{\left\{{
\begin{array}{l}
\alpha_t' =\alpha_t [6 \alpha_t + \alpha_b -16/3 \alpha_3- 3 \alpha_2-13/9 \alpha_1] \\
\alpha_b' =\alpha_b [6 \alpha_b + \alpha_t+\alpha_\tau -16/3 \alpha_3- 3 \alpha_2-7/9 \alpha_1] \\
\alpha_\tau' =\alpha_\tau [4 \alpha_\tau + 3 \alpha_b - 3 \alpha_2-3 \alpha_1] \\[2ex]
\alpha_c' =\alpha_c [3 \alpha_t -16/3 \alpha_3- 3 \alpha_2-13/9 \alpha_1] \\
\alpha_s' =\alpha_s [3 \alpha_b+\alpha_\tau -16/3 \alpha_3- 3 \alpha_2-7/9 \alpha_1] \\
\alpha_\mu' =\alpha_\mu [3 \alpha_b + \alpha_\tau- 3 \alpha_2-3 \alpha_1] \\[2ex]
A'=-A [\alpha_t + \alpha_b]/2 \\
\lambda'=0\\
\rho'=0 \\
\eta'=0\\[2ex]
M_{ij}'=M_{ij} [ \alpha_\tau (k_{i}+k_{j})/2 + 3 \alpha_t -3 \alpha_2-\alpha_1] 
\end{array}}
\right.}
$$
The symbol ' (=prime) denotes derivative 
with respect to $T=\log (Q/M_Z)/2 \pi.$
We define $\alpha_x={y_x^2}/{4 \pi}$ for $x=t,b,\tau,c,s,\mu,$
analogously to gauge $\alpha_i$'s.
$A,\lambda,\eta,\rho$ are the Wolfenstein parameters. 
$M_{ij}$ are the entries of neutrino mass matrix;
$k_3=1,$ and $k_2=0.$
$\alpha_1$ is normalized in standard 
model fashion--not SU(5)'s.

\end{document}